\documentclass[journal]{IEEEtran}
\ifCLASSINFOpdf
\else
\fi

\usepackage{amssymb}\usepackage[colorlinks]{hyperref}\usepackage{color}
\usepackage{stmaryrd}\usepackage{mathrsfs}
\usepackage{graphicx}
\usepackage{amsmath}
\usepackage{caption}

 \newcommand\emm[1]{{\ensuremath{#1}}}
\newcommand\prf{\textit{Proof. }}\newcommand\qed{\hfill\emm\blacksquare}

\newtheorem{thr}{Theorem} 
\newtheorem{lmm}{Lemma}
\newtheorem{cor}{Corollary}
 
\newtheorem{prp}{Proposition}

\newtheorem{dff}{Definition}

\newcommand\floor[1]{{\lfloor#1\rfloor}}

\newcommand{\lea}{<^+}
\newcommand{\gea}{>^+}
\newcommand{\eqa}{=^+}

\newcommand{\lel}{<^{\log}}
\newcommand{\gel}{>^{\log}}
\newcommand{\eql}{=^{\log}}

\newcommand{\lem}{\stackrel{\ast}{<}}
\newcommand{\gem}{\stackrel{\ast}{>}}
\newcommand{\eqm}{\stackrel{\ast}{=}}

\newcommand\m{{\mathbf {m}}}

\newcommand\K{{\mathbf K}} \newcommand\I{{\mathbf I}}

\renewcommand\d{{\mathbf d}} 
\newcommand\D{{\mathbf D}}

\newcommand\Ks{\Lambda} 
 
\newcommand\Q{{\mathbf Q}}
\newcommand\ch{{\mathcal H}} 
 
\newcommand\bb{{\mathbf {bb}}}

\newcommand\B{\mathbf{bb}}
 
\newcommand\R{\mathbb{R}}\renewcommand\Q{\mathbb{Q}}
\newcommand\N{\mathbb{N}}\newcommand\BT{\Sigma}
\newcommand\FS{\BT^*}\newcommand\IS{\BT^\infty}
\newcommand\FIS{\BT^{*\infty}}

\newcommand{\supp}{\mathrm{Supp}}

\renewcommand\qed{\hfill\emm\square}
\begin{document}
%
\title{All Sampling Methods Produce Outliers}
%
%
%

\author{Samuel Epstein
\thanks{JP Theory Group. samepst@jptheorygroup.org}}

\maketitle

\begin{abstract}
Given a computable probability measure $P$ over natural numbers or infinite binary sequences, there is no computable, randomized method that can produce an arbitrarily large sample such that none of its members are outliers of $P$. In addition, given a binary predicate $\gamma$, the length of the smallest program that computes a complete extension of $\gamma$ is less than the size of the domain of $\gamma$ plus the amount of information that $\gamma$ has with the halting sequence.
\end{abstract}


%
\IEEEpeerreviewmaketitle

\section{Introduction}
\label{sec:DefRand}

An outlier is a data point that varies noticeably from other data points in a sample or collection. There is no exact mathematical definition of what constitutes an outlier. Though there are known partial indicators, the determination of an outlier remains a subjective endeavor.

Outliers can have many causes, such as due to variability in system performance, human mistakes, instrument malfunctions, contamination from elements outside the population or by inherent standard deviations in populations.

In algorithmic information theory, outliers are precisely defined algorithmically with respect to computable probability measures over either natural numbers or infinite sequences. The probability measure represents the model, and natural numbers and infinite sequences are assumed to be data points with respect to these models. The level or score to which a data point is an outlier to a model (probability measure) is given by the \textit{deficiency of randomness} function. It is defined by $\d(x|P)=\floor{-\log P(x)}-\K(x|P)$, where $x$ is the data point and $P$ is the probability measure. The term $\K$ is the Kolmogorov complexity of a string, formally defined in Section \ref{sec:conv}. $\d(x|P)$ is the difference between length a string's $P$-code and its optimal description. If $x$ is not in the support of $P$, then $\d(x|P)=\infty$. The function $\d$ is optimal, in the following manner.

Given a computable probability measure $P$ over $\N$, an expectation bounded test is a function $d:\N\rightarrow\R_{\geq 0}$ that is lower semi-computable and
\begin{align*}
	\sum_{x\in\N}P(x)2^{d(x)}\leq 1.
\end{align*}
Lower semi-computability is formally defined in Section \ref{sec:conv}. Typical numbers $x$ of $P$ will have a low test score. An expectation bounded test $d$ is universal if for every expectation bounded test $d'$, there is a $c_{d'}\in\N$, such that for all $x\in\N$, $d(x)+c_{d'} > d'(x)$. 

It can be shown that the deficiency of randomness, $\d$, is a universal expectation test, in that there is a constant $c\in\N$, where for any expectation bounded test $d$, for any $x\in \N$,
\begin{align*}
	\d(x|P)+\K(d|P)&> d(x)-c.
\end{align*}

In this paper we show that all sampling methods produce outliers and provide a lower bound on the rate in which they occur. A sampling algorithm $A$ to a semi-measure $P$ is a computable function that takes as input a parameter $n\in\N$ and a random source of bits, and outputs, with probability one, an encoding of $2^n$ unique natural numbers. 

Note that if $P$ is a computable probability measure on $\N$, then for each $n\in\N$, there is only a finite number of $x\in\N$ where $\d(x|P)\leq n$. This is because there is an algorithm that on input $k\in\N$, can enumerate a list $M$ of numbers $x$ by order of $P(x)$ convergence time and stop when total $P$ mass of $M$ is $>1-2^{-k}$. $M$ is a finite set. Each $y$ in the support of $P$ and not in $M$ can be identified by a Shannon-Fano code of size  $\K(y|P)\lea-\log P(y)-k+\K(k)$ and thus has a deficiency of randomness $\gea k-\K(k)$. Thus any sampling method $A$ to a computable probability measure will, with increasing $n$ as input, produce samples containing members with increasing outlier scores,
\begin{align}
	\label{eq:lowboundsoutlier}
	\omega(1)&<\max_{a\in A(n)}\d(a|P).
\end{align}

For semi-measures in general, this bound is not necessarily guaranteed. For example the universal semi-computable semi-measure $\m$, defined in Section \ref{sec:conv}, has no outliers by definition. In this paper we improve the bounds of the above equation to a logarithmic scale, and prove the property holds for computable semi-measures.  \\

\noindent\textbf{Corollary.} 
\textit{For computable semi-measure $P$ over $\N$, for sampling method $A$, there is a constant $c_{P,A}\in\N$, such that for all $n,k\in \N$,
	$\Pr(n-\max_{a\in A(n)}\d(a|P)>k)<2^{-k+O(\K(k,n))+c_{P,A}}$}.\\

To achieve this result, we show that all sufficiently large sets will either have an outlier or high mutual information with the halting sequence. \\ \\
\noindent\textbf{Theorem.} 
\textit{Relativized to computable semi-measure $P$ over $\N$, for any finite set $D\subset\N$, $s = \floor{\log\sum_{a\in D}\m(a)/P(a)} < \log\max_{a\in D}\m(a)/P(a)+\I(D:\mathcal{H})+O(\K(s)+\K(\I(D:\mathcal{H})))$.}\\

The term $\I(D:\mathcal{H})=\K(D)-\K(D|\mathcal{H})$ is the  mutual information that $D$ has with the halting sequence. There is no computable method to produce sets $D$ that have arbitrary high mutual information with the halting sequence. We use this property to derive impossibility results of sampling methods. 

We also prove the same bounds with sampling methods over infinite sequences. The deficiency of randomness of an infinite sequence $\alpha\in\{0,1\}^\infty$ with respect to a computable probability measure $P$ over infinite sequences is $\D(\alpha|P)=\max_{n}-\log P(\alpha[1..n])-\K(\alpha[1..n]|P)$. If $\alpha$ is not in the support of $P$, then $\D(\alpha|P)=\infty$. $\D$ is universal over integral tests (see Section \ref{sec:DefInfSec}). A continuous sampling method $A$ to a probability measure $P$ takes in a parameter $n$ and an infinite source of random bits and outputs $2^n$ unique infinite sequences, encoded in the form $\alpha_1[1]\alpha_2[1]...\alpha_{2^n}[1]\alpha_1[2]\alpha_2[2]...\alpha_{2^n}[2]\dots$ 
We get the following sampling corollary which is analogous to the discrete case.\\

\noindent\textbf{Corollary.}
\textit{For computable measure $P$ over $\{0,1\}^\infty$, for continuous sampling method $A$, there is a constant $c_{P,A}\in\N$, such that for all $n,k\in \N$,
	$\Pr(n-\max_{\alpha\in A(n)}\D(\alpha|P)>k)<2^{-k+O(\log k+\K(n))+c_{P,A}}$.} \\ 

This theorem was derived similarly to the discrete case, by first showing that large sets of infinite sequences with low $\D$ scores have high information with the halting sequence. The information term $\I$ over infinite sequences used in this paper was introduced in \cite{Levin74}. The continuous sampling no-go corollary is derived from the following theorem, similarly to the discrete case. The term $\langle Z\rangle$ is defined in Section \ref{sec:conv}.\\

\noindent\textbf{Theorem.}
\textit{	Relativized to computable probability measure $P$ over $\{0,1\}^\infty$, for any $Z\subseteq \{0,1\}^\infty$, if $\N\ni s< \log\sum_{\alpha\in Z}2^{\D(\alpha|P)}$, then $s<\sup_{\alpha\in Z}\D(\alpha|P)\,{+}\,\I(\langle Z\rangle:\ch)+O(\K(s)+\log\I(\langle Z\rangle:\ch))$.}
\subsection{Binary Predicates}

In this paper, we also prove upper bounds on the size of the smallest program that computes a complete extension of a given binary predicate $\gamma$. We prove that for non-exotic predicates, this size is not more than the number of elements of $\gamma$. Exotic predicates have high mutual information with the halting sequence, and thus no algorithm can generate such predicates. 

More formally, a binary predicate is defined to be a function of the form $f:D\rightarrow\{0,1\}$, where $D\subseteq\N$.  We say that binary predicate $\lambda$ is an extension of $\gamma$, if  for all $i\in \mathrm{Dom}(\gamma)$, $\gamma(i)=\lambda(i)$. If a binary predicate has a domain of $\N$ and is an extension of binary predicate $\gamma$, then we say it is a complete extension of $\gamma$. In this paper we prove the following result.\\

\noindent\textbf{Theorem.}
\textit{For binary predicate $\gamma$ and the set $\Gamma$ of complete extensions of $\gamma$, $\min_{g\,\in\,\Gamma}\K(g)\,{\lel}\,|\mathrm{Dom}(\gamma)|{+}\I(\langle\gamma\rangle{:}\mathcal{H}).$}

\section{Related Work}

The study of Kolmogorov complexity originated from the work of~\cite{Kolmogorov65}. The canonical self-delimiting form of Kolmogorov complexity was introduced in~\cite{ZvonkinLe70} and treated later in~\cite{Chaitin75}. The universal probability $\m$ was introduced in~\cite{Solomonoff64}. More information about the history of the concepts used in this paper can be found the textbook~\cite{LiVi08}. 

Information conservation laws were introduced and studied in~\cite{Levin74,Levin84}. Information asymmetry and the complexity of complexity were studied in~\cite{Gacs74}.  A history of the origin of the mutual information of a string with the halting sequence can be found in~\cite{VereshchaginVi04v2}.

The notion of the deficiency of randomness with respect to a measure follows from the work of~\cite{Shen83}, and also studied in~\cite{KolmogorovUs87,Vyugin87,Shen99}. At a Tallinn conference in 1973, Kolmogorov formulated the notion of a two part code and introduced the structure function (see~\cite{VereshchaginVi04v2} for more details). Related aspects involving stochastic objects were studied in~\cite{Shen83,Shen99,Vyugin87,Vyugin99}. 

The combination of complexity with distortion balls can be seen in~\cite{FortnowLeVe06}. The work of Kolmogorov and the modeling of individual strings using a two-part code was expanded upon in~\cite{VereshchaginVi04v2,GacsTrVi01}.  These works introduced the notion of  using the prefix of a ``border'' sequence to define a universal algorithmic sufficient statistic of strings. The generalization and synthesis of this work and the development of algorithmic rate distortion theory can be seen in the works of~\cite{VereshchaginVi04,VereshchaginVi10}. More information on algorithmic statistics can be found in \cite{VereshchaginSh17,VereshchaginSh15}.

The outlier theorem is an extension to the ``Sets Have Simple Members" theorem, first appearing in \cite{EpsteinLe11}. This theorem was derived from the work in~\cite{EpsteinBe2011}, which introduced a variant of Theorem 6 in~\cite{VereshchaginVi04}. The first game theoretic proof to the ``Sets Have Simple Members'' theorem can be found in~\cite{Shen12}.

The formulas in this paper involving information with the halting sequence are compatible with the Independence Postulate, detailed in \cite{Levin84,Levin13}. The Independence Postulate is a generalization of the Church-Turing thesis.

\section{Conventions}
\label{sec:conv}
We use $\N$, $\mathbb{Z}$, $\Q$, $\R$, $\BT$, $\FS$, and $\IS$ to represent natural numbers, integers, rational numbers, reals, bits, finite strings, and infinite strings. Let $X_{\geq 0}$ and $X_{>0}$ be the sets of non-negative and of positive elements of $X$. The length of a string $x{\in}\BT^n$ is denoted by $\|x\|=n$. The removal of the last bit of a string is denoted by $(p0^-){=}(p1^-){=}p$, for $p\in\FS$. For the empty string $\emptyset$, $(\emptyset^-)$ is undefined. We use $\FIS$ to denote $\FS{\cup}\IS$, the set of finite and infinite strings.  For $x\in \FIS$, $y\in \FIS$, we say $x\sqsubseteq y$ if $x=y$ or $x\in\FS$ and $y=xz$ for some $z\in \FIS$. Also $x\sqsubset y$ if $x\sqsubseteq y$ and $x\neq y$. The $i$th bit of a string $x\in\FIS$ is denoted by $x[i]$. The first $n$ bits of a string $x\in\FIS$ is denoted by $x[0..n]$. The indicator function of a mathematical statement $A$ is denoted by $[A]$, where if $A$ is true then $[A]=1$, otherwise $[A]=0$. The size of a finite set $S$ is denoted to be $|S|$. We use $\langle x\rangle$ to represent a self delimiting code for $x\in\FS$, such as $1^{\|x\|}0x$. The self delimiting code for a finite set of strings $\{a_1,\dots,a_n\}$ is $\langle\{a_1,\dots,a_n\}\rangle=\langle n\rangle\langle a_1\rangle\langle a_2\rangle\dots\langle a_n\rangle$.  For two infinite strings $\alpha$ and $\beta$, $\langle \alpha,\beta\rangle=\alpha_1\beta_1\alpha_2\beta_2\dots$. For sets $Z$ of infinite strings, $Z_{\leq n}=\{\alpha[0..n]\,{:}\,\alpha\,{\in}\,Z\}$ and $\langle Z\rangle=\langle Z_{\leq 1}\rangle\langle Z_{\leq 2}\rangle\langle Z_{\leq 3}\rangle\dots$. 

As is typical of the field of algorithmic information theory, the theorems in this paper are relative to a fixed universal  machine, and therefore their statements are only relative up to additive and logarithmic precision. For positive real functions $f$ the terms  ${\lea}f$, ${\gea}f$, ${\eqa}f$ represent ${<}f{+}O(1)$, ${>}f{-}O(1)$, and ${=}f{\pm}O(1)$, respectively. In addition ${\lem}f$, ${\gem}f$ denote $<f/O(1)$, $>f/O(1)$. The terms ${\eqm}f$  denotes ${\lem}f$ and ${\gem}f$. For nonnegative real function $f$, the terms ${\lel}f$, ${\gel} f$, ${\eql}f$ represent the terms ${<}f{+}O(\log(f{+}1))$, ${>}f{-}O(\log(f{+}1))$, and ${=}f{\pm}O(\log(f{+}1))$, respectively. A discrete measure is a nonnegative function $Q:\N\rightarrow \R_{\geq 0}$ over natural numbers. The support of a measure $Q$ is the set of all elements whose $Q$ value is positive, with $\supp(Q) = \{a\,{:}\,Q(a)>0\}$. A measure is elementary if its support is finite and its range is a subset of $\Q$. We say $Q$ is a semi-measure if $\sum_aQ(a)\,{\leq}\,1$. We say that $Q$ is probability measure if $\sum_aQ(a)\,{=}\,1$.

$T_y(x)$ is the output of algorithm $T$ (or $\perp$ if it does not halt) on input $x\in\FS$ and auxiliary input $y\in\FIS$. $T$ is prefix-free if for all $x,s\in\FS$ with $s\,{\neq}\,\emptyset$, and $y\in\FIS$, either $T_y(x)\,{=}\perp$ or $T_y(xs)\,{=}\perp$ . The complexity of $x\in\FS$ with respect to $T_y$ is $\K_T(x|y)= \min\{\|p\|\,:\,T_y(p)=x\}$. 

There exists optimal for $\K$ prefix-free algorithm $U$, meaning that for all prefix-free algorithms $T$,  there exists $c_T\,{\in}\,\N$, where $\K_U(x|y)\leq \K_T(x|y)+c_T$ for all $x\,{\in}\,\FS$ and $y\,{\in}\,\FIS$. For example, one can take a universal prefix-free algorithm $U$, where for each prefix-free algorithm $T$, there exists  $t\in\FS$, with $U_y(tx)=T_y(x)$ for all $x\in\FS$ and $y\in\FIS$. The function $\K(x|y)$, defined to be $\K_U(x|y)$, is the Kolmogorov complexity of $x\in\FS$ relative to $y\in\FIS$. When we say that a universal Turing machine is relativized to an object, this means that an encoding of the object is provided to the universal Turing machine on an auxiliary tape.

A function $f:\N\rightarrow\R$ is computable if there is a total recursive function $g(x,n)$ over all $x\in\N$ and $n\in\N$ where $|f(x)-g(x,n)|<1/n$.  The complexity of such a computable function $f$, is $\K(f)$, the minimal length of a $U$-program to compute $f$. A function $f:\N\rightarrow\R$ is lower semi-computable if the set $S=\{(x,r):x\in\N, r\in Q, r<f(x)\}$ is recursively enumerable. If $f$ is not computable but lower semi-computable, then its complexity $\K(f)$ is equal to the size of smallest $U$-program that on input $x$, enumerates $\{r:f(x)> r\}$.

The chain rule for Kolmogorov complexity is $\K(x,y) \eqa \K(x)+\K(y|\langle x,\K(x)\rangle)$.  The mutual information in finite strings $x$ and $y$ relative to $z\in\FS$ is $\I(x\,{:}\,y\,{|}\,z)= \K(x|z)+\K(y|z)-\K(\langle x,y\rangle|z)\eqa\K(x|z)-\K(x|\langle y,\K(y|z),z\rangle)$.  The universal probability of a number $a\in\N$ is $\m(a|y){= }\sum_z[\,U_y(z)=a]2^{-\|z\|}$. The coding theorem states $-\log \m(a|y)\eqa\K(a|y)$.

The halting sequence $\mathcal{H}\in\IS$ is the infinite string where $\mathcal{H}[i]=[U(i)\neq \perp]$ for all $i\in\N$. As mentioned in the introduction, the amount of information that $a\in\N$ has with $\mathcal{H}$ is denoted by $\I(a:\mathcal{H})=\K(a)-\K(a|\mathcal{H})$.

\section{Algorithmic Statistics}
Algorithmic Statistics is the study of the separation of information, i.e. a string $x\in\FS$, into two parts. The first part is the model containing the ``denoised” information of $x$. The second part is the data-to-model code representing the remaining randomness in $x$. The algorithmic statistics that we use in this paper are computable semi-measures $P$ which have $x$ in their support. Other models studied in the literature are finite setspf numbers and total recursive functions. For semi-measures, the model is an encoding or Turing number of an algorithm that computes $P$. The data-to-model code is the Shannon Fano encoding of length $\eqa-\log P(x)$ of $x$ with respect to $P$. If $x$ is typical of a model then it has a low deficiency of randomness $\d(x|P)=\floor{-\log P(x)}-\K(x|P)$.

The field of algorithmic statistics studies properties of algorithmic sufficient statistics, i.e. statistics whose sum of the model complexity and data-to-model code length is equal (up to a small error term) to $\K(x)$. For probability distributions, these are such $P$ where $\K(P)-\log P(x) \approx \K(x)$. A minimal sufficient statistic is an algorithmic sufficient statistic with the smallest model complexity, i.e. one that minimizes $\K(P)$. According to Occam’s razor, out of all the algorithmic sufficient statistics, the minimal ones summarize the relevant information of x in the most concise manner.

This paper is connected to algorithmic statistics in two ways. First, the main theorem is a result about deficiencies of randomness, $\d$. The deficiency function $\d$ and its relation to models are one of the central areas of study in algorithmic statistics. Second,  Lemma \ref{lmm:main} is a statement about the stochasticity measure of a finite set of strings. The stochasticity term is related to those used in algorithmic statistics in that it measures whether a string is typical of a simple probability measure. The extended deficiency of randomness of $x$ with respect to elementary measure $Q$ and $v\in\N$ is $\d(x|Q,v)=\floor{-\log Q(x)}-\K(x|\langle Q\rangle,v)$. The stochasticity of $a\in\N$, conditional to $b\in\N$, is measured by\\
\begin{dff}[Stochasticity]\\
	$\Ks(a|b)=\min\{\K(Q|b)+3\log\max\{\d(a|Q,b),1\}\,{:}$\\ $\,\textrm{$Q$ is an elementary probability measure}\}.$\\
\end{dff}
We have $\Ks(a)=\Ks(a|\emptyset)$, with $\Ks(a|b)\lea\Ks(a)+\K(b)$.
Thus if $a$ has low $\Ks(a)$, then it is typical for a simple probability measure. Stochasticity is an important area of research because the stochasticity measure of an elementary object lower bounds the amount of information that the object has with the halting sequence, as shown in Section \ref{sec:LeftTotal}. Objects with high mutual information with the halting sequence are exotic in that there is no (randomized) method to produce them, due to information nongrowth laws. Thus the study of stochasticity yields insight into the properties of objects that can and cannot be produced by algorithms.

\section{Games}
\label{sec:games}

In this section we introduce a generalization to the so-called ``Epstein-Levin'' game, introduced in \cite{Shen12}. This new generalized game consists of a finite bipartite graph $E\subseteq L\times R$, with $L\subset \N$ and $R\subset \N$. There is a computable probability distribution $P$ over the right vertices. The game is between Alice and Bob and is defined by four additional parameters.
\begin{enumerate}
	\item An integer $k$.
	\item A positive rational $l$.
	\item A positive rational number $\delta$.
	\item A computable function $W:\N\rightarrow\R_{\geq 0}$.
\end{enumerate}
The rules of the game are as follows. Alice assigns increasing rational nonnegative \textit{weights} to vertices on $L$, which are all initially 0. The sum $\sum_{a\in L}W(a)\cdot\mathrm{weight}(a)$ cannot exceed 1. After each turn by Alice, Bob can mark vertices on $L$ and $R$. Once a vertex is marked, it will stay marked. There are restrictions on how Bob can mark the left and right vertices. The sum $W(a)$ over all marked left vertices cannot exceed $l$. Furthermore, the total $P$-probability of marked vertices on the right is at most $\delta$.

Bob wins if every $R$ vertex whose combined weight of its $L$-neighbors is equal to or greater than $2^{-k}$ either has a marked neighbor or is marked itself. Note that this is a generalization of the ``Epstein-Levin'' game in \cite{Shen12}, whose instantiation is equivalent to setting $W(a)=1$ for all $a\in\N$.

\begin{lmm}
	\label{lmm:BobWin}
	For $l=O(2^k\log(1/\delta))$, Bob has a computable winning strategy.
\end{lmm}

Note that the game can be made finite by making the weights restricted to the form $2^{-m}$ for $m\in\mathbb{Z}$. Since this new game changes the weights by a factor of at most 2, Bob can compensate by changing $k$ by 1. In addition, the minimal weight is changed to be an $m\in\mathbb{Z}$ where $2^{-m}\max_{a\in L}W(a)|L|<1$ so the sum $\sum_{a\in L}W(a)\cdot\mathrm{weight}(a)\leq 2$, which is a constant factor. Thus this game is a finite game with full information so either Alice or Bob has a winning strategy. We prove that Bob has a probabilistic strategy that has a non-zero chance of winning. Thus Alice can't have a winning strategy, otherwise Bob's strategy would succeed with probability 0. Bob's simple probabilistic strategy is unchanged from that in \cite{Shen12}:
\begin{itemize}
	\item If Alice increases the weight on a vertex $a\in L$, by some value $\varepsilon\in(0,1]$, then Bob marks that vertex with probability $c2^k\varepsilon$, where $c>1$ is a constant to be chosen later. If $c2^k\varepsilon>1$, then Bob marks the vertex.
	\item If a vertex on $R$ has neighbors in $L$ with total weight not less than $2^{-k}$ but no marked neighbors, then Bob immediately marks this vertex.
\end{itemize}
To prove that Bob has a non-zero chance of succeeding, we prove the following two events each have probability less than 1/2.
\begin{enumerate}
	\item \label{HEvent1}The total $P$-measure of marked $R$-vertices exceeds $\delta$.
	\item \label{HEvent2}The sum of $W(a)$ over all marked left vertices $a\in L$ is more than $l$.
\end{enumerate}
For (\ref{HEvent1}), for each $y\in R$, with left neighbors with weights increasing $\varepsilon_1,\dots,\varepsilon_m$, with $\sum\varepsilon_i\geq 2^{-k}$, the probability that all its neighbors are unmarked is not more than
$$
(1-c2^k\varepsilon_1)\dots(1-c2^k\varepsilon_m)\leq e^{-c2^k(\varepsilon_1+\dots+\varepsilon_m)}\leq e^{-c}.
$$
For every $P$-measure, the expected $P$-measure of marked vertices in $R$ does not exceed $e^{-c}$. For (\ref{HEvent1}) to be less than 1/2, it suffices for $c=\ln(1/\delta)+O(1)$.

For (\ref{HEvent2}), the requirement that $\sum_{a\in L}\label{key}W(a)\cdot\mathrm{weight}(a)\leq 1$ guarantees the following bound on the expectation
\begin{align*}
&E\left[\sum\left\{W(a):\textrm{$a$ is a left marked vertex}\right\}\right]\\
&\leq \sum_{a\in L}W(a)\cdot\mathrm{weight}(a)c2^k\\
&\leq c2^k.
\end{align*}
Thus (\ref{HEvent2}) is satisfied for $l=c2^{k+2}=O(2^k\log(1/\delta))$, thus proving the lemma.

\subsection{Stochasticity}
The above game can be applied to the following statement about the stochasticity of finite sets of natural numbers.
\begin{lmm}
	\label{lmm:main}
	Let $\eta:\N\rightarrow \R_{\geq 0}$ be a lower semi-computable function, $W:\N\rightarrow \R_{\geq 0}$ be a computable function with $\sum_{a\in\N}W(a)\eta(a)\leq 1$. Then for every finite $D\subset \N$ with $\log\sum_{a\in D}\eta(a)\geq s\in\mathbb{Z}$ there is $a\in D$ with $\K(a)\lea -\log W(a)-s+\Lambda(D)+2\K(s)$. Note the above is true relative to any oracle $\alpha$.
\end{lmm}

\prf
	Let $Q$ be any elementary probability measure witnessing $\Ks(\langle D\rangle|s)$. The  randomness deficiency of $\langle D\rangle$ with respect to $Q$,  conditional to $s$, is $d=\max\{\d(\langle D\rangle|Q,s),1\}$. From $Q$ we create the following generalized Epstein-Levin game. The bipartite graph $E\subseteq L\times R$ is created by having $R$ be the encoded sets $\langle G\rangle$ in the support of $Q$. The combined members of encoded sets in $R$ are set to $L$ and there is a connection between a vertex $a\in L$ and an encoded set $\langle G\rangle\,{\in}\, R$, if and only if $a\in G$. Alice approximates the weights $\eta$ from below. At each round, Alice increases the weight of a vertex in $L$ by the amount specified in the corresponding round of the lower enumeration of $\eta$. We set the parameters $k=-s$ and $\delta = 2^{-cd}$, for a constant $c\in\N$ solely dependent on the universal Turing machine to be determined later. The elementary probability is $P=Q$. By Lemma \ref{lmm:BobWin},  Bob has a winning strategy where the sum of all $W(a)$ over left vertices marked by Bob is at most
	$$
	l=O(2^k\log(1/\delta))=O(cd2^{-s}).
	$$
	The right vertex $\langle D\rangle$ is not marked. Otherwise, since the $Q$ measure of vertices that are marked is not more than $2^{-cd}$, and right vertices are marked during the course of the game, the function $Q'=(Q\cdot 2^{cd})$ restricted to marked right vertices is a lower semi-computable semi-measure. This semi-measure can be lower computed using $Q$, $d$, $s$, and $c$. Hence the $Q'$ code of $D$ would have the size $\eqa -\log(Q(\langle D\rangle)2^{cd})$. Thus the following contradiction occurs for large enough $c\in\N$ dependent solely on the universal Turing machine $U$,
	\begin{align*}
	&\;\;\;\;\;\;\K(\langle D\rangle |\langle Q\rangle,d,s,c)\\
	&\lea -\log Q(\langle D\rangle)-cd\\
	cd &\lea -\log Q(\langle D\rangle)-\K(\langle D\rangle|\langle Q\rangle,s)+\K(c,d)\\
	cd &\lea d+\K(c,d).
	\end{align*}
	Therefore, since $\langle D\rangle$ is not marked, and since $\sum_{a\in D}\eta(a)\geq 2^{s}=2^{-k}$, by the rules of the game, $D$ has a marked $a\in L$. The semi-measure $p(a)=W(a)/l$ for Bob's marked $L$ vertices is lower semi-computable relative to $Q$, $s$, and $d$, so
	\begin{align*}
	&\;\;\;\;\;\;\K(a|Q,s,d) \\
	& \lea -\log p(a)\\
	&\lea -\log W(a)+\log l\\
	&\lea -\log W(a)-s+\log d\\
	\K(a)&\lea -\log W(a)-s+\K(d)+\log d+\K(Q|s)+\K(s)\\
	\K(a)&\lea -\log W(a)-s+\Lambda(\langle D\rangle|s)+\K(s)\\
	\K(a)&\lea -\log W(a)-s+\Lambda(\langle D\rangle)+2\K(s).
	\end{align*} \qed

\subsection{Stochastic Sets}
The above lemma can be applied to the following result showing that large sets of numbers with low randomness deficiencies are exotic.\\

\begin{thr}
	\label{thr:stochsamp}	
	Relativized to computable semi-measure $P$ over $\N$, for any finite set $D\subset \N$, if $\N\ni s <\log\sum_{a\in D}\m(a)/P(a)$, then $s\lea\log \max_{a\in D}\m(a)/P(a)+\Lambda(D)+2\K(s)$.\\
\end{thr}

\prf
	We invoke Lemma \ref{lmm:main}. $W(a)$ is set to $P(a)$. $\eta(a)$ is set to $\m(a)/P(a)$ and is thus lower semi-computable. In addition $\sum_{a\in\N}W(a)\eta(a)\leq 1$ and $\sum_{a\in D}\eta(a)\geq 2^s$.  The lemma produces an $a\in\N$ such that  $\K(a)\lea -\log P(a) -s +\Lambda(D)+2\K(s)$. Some reworking proves the theorem.\qed\\

\begin{cor}
	\label{cor:stochsamp}	
	Relativized to computable semi-measure $P$ over $\N$, for any finite set $D\subset \N$, if $\N\ni s <\log |D|$, then $s\lea\log \max_{a\in D}\m(a)/P(a)+\Lambda(D)+2\K(s)$.
\end{cor}

\section{Helper Lemmas}
The following elementary lemmas are used, in a helper capacity, throughout the paper. The terminology $O(f)$ for some function $f:\N\rightarrow\N$ signifies Big Oh notation of $f$ with the parameters solely dependent on the choice of the universal Turing machine $U$. This \ holds also for the $f\lea g$ inequality, which is equal to $f<g+O(1)$, for functions $f,g$ between $\N$.\\
\begin{lmm}
	\label{lmm:monotonic}
	For every $c,n\in\N$ there exists $c'\in\N$ where if $x<y+c$ for some $x,y\in\N$ then $x+n\K(x)< y+n\K(y)+c'$.\\
\end{lmm}
\prf
	$\K(x)\lea \K(y)+\K(y-x)$ as $x$ can be computed from $y$ and $(y-x)$. So $n\K(x)-n\K(y) < n\K(y-x)+O(n)$. Assume not, then there exists $x,y,c\in\N$ where $x<y+c$
	and $y-x+c'< n\K(x)-n\K(y)<n\K(y-x)+O(n)$, which is a contradiction for $c'=O(n)+2c+\max_a
	\{2n\log a-a\}$.\qed\\

\begin{lmm}
	\label{lmm:abc}
	For $d,d',n,m\in\N$ there exists $d''\in \N$ where for any $f,g,h\in\N$, if $f<g+n\K(g)+d$ and $g<h+m\K(h)+d'$, then $f<h+(m+2n)\K(h)+d''$.\\
\end{lmm}
\prf
	If $g< h+d'$, then due to Lemma \ref{lmm:monotonic} applied to $x=g$, $y=h$, $n$, and $c=d'$, there exists $c'$ dependent on $d'$ and $n$ where $g+n\K(g)<h+n\K(h)+c'$ and thus  $f< h +n\K(h)+d+c'$, proving the lemma. Thus $h+d'\leq g$ and $g-h< m\K(h)+d'$, which implies $\K(g-h)\lea 2\log m\K(h)+2
	\log d'$. Therefore $\K(g)\lea\K(h)+\K(g-h)\lea \K(h)+2\log m+2\log \K(h)+2\log d'$. So,
	\begin{align*}
	f & < g+ n\K(g)+d\\
	&\lea h+m\K(h										)+n\K(g)+d+d'\\
	&\lea h+m\K(h)+n(\K(h)+\K(g-h)+O(1))+d+d'\\
	&\lea h+m\K(h)+n(\K(h)+2\log m\\
	&\;\;\;\;\;\;+2\log \K(h)+2\log d'+O(1))+d+d'\\
	&<h+(m+2n)\K(h)+O(n\log m)+2n\log d' +d+d'\\
	&<h+(m+2n)\K(h)+d'',\
	\end{align*}
	where $d''=O(n\log m)+2n\log d' +d+d'$.\qed\\

\begin{lmm}
	\label{lmm:log} For every $c,n\in \N$, there exists $c'\in\N$ where for all $x,y\in\N$, if $x<y+n\log x+c$ then $x<y+2n\log y +c'$.\\
\end{lmm}
\prf
	\begin{align*}
	\log x &< \log y + \log \log x+\log cn\\
	2\log x - 2\log \log x &<2\log y+2\log cn\\
	\log x &< 2 \log y + 2 \log cn.
	\end{align*}
	Combining with the original inequality 
	\begin{align*}
	x &< y + n \log x+c\\
	x &< y+n(2\log y +2\log cn)+c\\
	&=y+2n\log y +c',
	\end{align*} 
	where $c'=2n\log cn +c$.\qed\\

\begin{lmm}
	\label{lmm:K} For every $d\in\N$ there exists $d'\in\N$ where if $x<y+\K(x)+d$ then $x<y+2\K(y)+d'$.\\
\end{lmm}
\prf
	It must be that $y+d< x$, otherwise the lemma is trivially solved. Thus $x-y< \K(x)+d$, so $\K(x-y) \lea 2\log \K(x)+2\log d$. So $\K(x) \lea \K(y) + \K(x-y) \lea \K(y) + 2\log \K(x)+2\log d$. By Lemma \ref{lmm:log}, where $n=2$ and $c=2\log d+O(1)$,  there is a $c'\in\N$, where $\K(x) < \K(y) + 4 \log \K(y)+c'< 2\K(y)+c'+O(1)$. So 
	\begin{align*}
	x&<y+\K(x)+d\\
	&<y+2\K(y)+c'+d+O(1)\\
	&=y+2\K(y)+d',
	\end{align*}
	where $d'=c'+d+O(1)$. \qed\\

\begin{lmm}
	\label{lmm:aKgb} 
	For all $d,m\in\N$ there is a $d'\in\N$ where if $x+m\K(x)+d>y$ then $x+d'>y-2m\K(y)$.\\
\end{lmm}
\prf
	If $x+d>y$, then the lemma is satisfied, so $x+d\leq y$. Thus $y-x< m\K(x)+d$ implies $\K(y-x)\lea 2\log \K(x)+2\log dm$. Thus $\K(x)\lea \K(y)+\K(y-x)\lea \K(y)+2\log \K(x)+2\log dm$. Applying Lemma \ref{lmm:log} where $c=2\log dm+O(1)$ and $n=2$, we get a $c'$ dependent on $c$ and $n$ where $\K(x)< \K(y)+4\log \K(y)+c'<2\K(y)+c'+O(1)$. So
	\begin{align*}
	x+m\K(x)+d&>y\\
	x+m(2\K(y)+c'+O(1))+d&>y\\
	x+d'&>y-2m\K(y),
	\end{align*}
	where $d' = m(c'+O(1))+d$.
	\qed
\section{Left-Total Machines}
\label{sec:LeftTotal}
\begin{figure}[h!]
	\begin{center}
		\includegraphics[width=0.7\columnwidth]{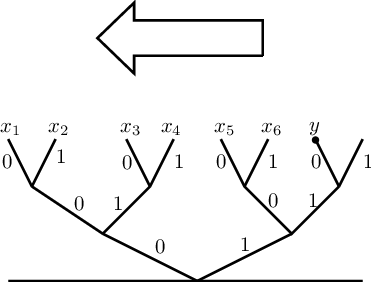}
		\caption{The above diagram represents the domain of a left total machine $T$ with the 0 bits branching to the left and the 1 bits branching to the right. For $i\in \{1,\dots,5\}$, $x_i\lhd x_{i+1}$ and $x_i\lhd y$. Assuming $T(y)$ halts, each $x_i$ is total. This also implies each $x_i^-$ is total as well.}
		\label{fig:LeftTotal}
	\end{center}
\end{figure}

We say $x\in\FS$ is total with respect to a machine if the machine halts on all sufficiently long extensions of $x$. More formally, $x$ is total with respect to $T_y$ for some $y\in\FIS$ if there exists a finite prefix free set of strings $Z\subset\FS$ where $\sum_{z\in Z}2^{-\|z\|}=1$ and $T_y(xz)\neq\perp$ for all $z\in Z$.  We say $\alpha\in\FIS$ is to the ``left'' of $\beta\in\FIS$, and use the notation $\alpha\lhd \beta$, if there exists $x\in\FS$ such that $x0\,{\sqsubseteq}\, \alpha$ and $x1\,{\sqsubseteq}\, \beta$. A machine $T$ is left-total if for all auxiliary strings $\alpha\in\FIS$ and for all $x,y\in\FS$ with $x\lhd y$, one has that $T_\alpha(y)\neq\perp$ implies that $x$ is total with respect to $T_\alpha$. An example  left-total machine can be seen in Figure \ref{fig:LeftTotal}.

For the remaining part of this paper, we can and will change the universal self delimiting machine $U$ into an optimal left-total machine $U'$ by the following definition. The algorithm $U'$ enumerates all strings $p\,{\in}\,\FS$ in order of their convergence time of $U(p)$ and successively assigns them consecutive intervals $i_p{\subset}[0,1]$ of width $2^{-\|p\|}$. Then $U'$ outputs $U(p)$ on input $p'$ if the open interval corresponding to $p'$ and not that of $(p')^{-}$ is strictly contained in $i_p$. The open interval in [0,1] corresponding with $p'$ is $((p')2^{-\|p'\|},((p'){+}1)2^{-\|p'\|})$ where $(p)$ is the value of $p$ in binary. For example, the value of both strings 011 and 0011 is 3. The value of 0100 is 4. The same definition applies for the machines $U'_\alpha$ and $U_\alpha$, over all $\alpha\,{\in}\,\FIS$. We now set $U$ to equal $U'$.

\begin{figure}[h!]
	\begin{center}
		\includegraphics[width=0.7\columnwidth]{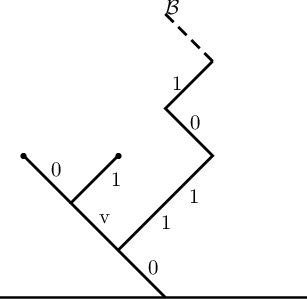}
		\caption{The above diagram represents the domain of the optimal left-total algorithm $U$, with the 0 bits branching to the left and the 1 bits branching to the right. The strings in the above diagram, $0v0$ and $0v1$, are halting inputs to $U$ with $U(0v0)\neq \perp$ and $U(0v1)\neq \perp$. So $0v$ is a total string. The infinite border sequence $\mathcal{B}\in\IS$ represents the unique infinite sequence such that all its finite prefixes have total and non total extensions. All finite strings branching to the right of $\mathcal{B}$ will cause $U$ to diverge.}
		\label{fig:DomainUPrime}
	\end{center}
\end{figure}

Without loss of generality, the complexity terms of this paper are defined with respect to the optimal left total machine $U$. The infinite border sequence $\mathcal{B}\in\IS$ represents the unique infinite sequence such that all its finite prefixes have total and non total extensions. The term ``border'' is used because for any string $x\in\FS$, $x\lhd\mathcal{B}$ implies that $x$ total with respect to $U$ and $\mathcal{B}\lhd x$ implies that $U$ will never halt when given $x$ as an initial input. Figure~\ref{fig:DomainUPrime} shows the domain of $U$ with respect to $\mathcal{B}$. 

\subsection{Properties of Total Strings}
This section uses the notion of a Martin L\"{o}f random infinite sequence. An infinite sequence $\alpha\in\IS$ is Martin L\"{o}f random if there is a constant $c\in\N$ such that for all $n\in\N$, $\K(\alpha[0..n])> n-c$. Let $\Omega =\sum_x\m(x)$ be Chaitin's Omega, the probability that U will halt. It is well known that the binary expansion of $\Omega$ is Martin L\"{o}f random. \\
\begin{prp}
	\label{prp:borderprefix}
	The border sequence $\mathcal{B}$ is Martin L\"{o}f random. Furthermore if $b\in\FS$ is total and $b^-$ is not, then $b^-\sqsubset\mathcal{B}$.\\
\end{prp}
\prf
	The border sequence is the binary expansion  of Chaitin's Omega for machine U, because the probability that a random infinite sequence contains a prefix that is a halting program is precisely the probability that the random sequence is at the left of the border sequence. If $b\in \FS$ is total and $b^-$ is not, then $b^-$ has a total extension $b^-0$ and a non total extension $b^-1$, thus by the definition of the border sequence, $b^-\sqsubset\mathcal{B}$.\qed\\
\begin{lmm}
	\label{lmm:totalString}
	If $b\in\FS$ is total and $b^-$ is not, and $x\in\FS$, \\
	then $\K(b)+\I(x:\mathcal{H}|b)\lel \I(x\,{:}\,\mathcal{H})+\K(b|\langle x,\|b\|\rangle)$.\\
\end{lmm}
\prf
	By Proposition \ref{prp:borderprefix}, $b^-\sqsubset\mathcal{B}$ is a prefix of the border sequence and thus $\|b\|\lea\K(b)$. Since $\mathcal{B}$ is computable from the halting sequence $\mathcal{H}$, we have that $b$ is computable from $\|b\|$ and $\mathcal{H}$, with $\K(b|\mathcal{H})\lea \K(\|b\|)$. 
	
	The chain rule gives the equality $\K(b)+\K(x|b,\K(b)) \eqa \K(x)+\K(b|x,\K(x))$. Combined with the inequalities $\K(x|b)\lea \K(x|b,\K(b))+\K(\K(b))$ and $\K(b|x,\K(x))\lea \K(b|x)$, we get
	\begin{align*}
	\K(b)+\K(x|b)&\lea \K(x)+\K(b|x)+\K(\K(b)).
	\end{align*} 
	
	Subtracting $\K(x|b,\mathcal{H})$ from both sides results in
	\begin{align*}
	&\;\;\;\;\;\;\K(b)+\K(x|b)-\K(x| b,\mathcal{H})\\
	&\lea \K(x)+\K(b|x)+\K(\K(b))-\K(x| b,\mathcal{H})\\
	&\lea \K(x)+\K(b|x)+\K(\K(b))-\K(x|\mathcal{H})+\K(b|\mathcal{H}).\\
	&\lea\I(x:\mathcal{H})+\K(b|x)+\K(\K(b))+\K(b|\mathcal{H})\\
	&< \I(x:\mathcal{H})+\K(b|x)+O(\log\|b\|)\\
	&<\I(x:\mathcal{H})+ \K(b|\langle x,\|b\|\rangle)+\K(\|b\|)+O(\log \|b\|).
	\end{align*}
	So $ \K(b)+\I(x:\mathcal{H}|b)\lel \I(x:\mathcal{H})+\K(b|\langle x,\|b\|\rangle)$.\qed\\
\begin{lmm}
	\label{lmm:KTotalLength}
	If $b\in\FS$ is total and $b^-$ is not, and for $x\in \FS$, $\K(b|\langle x,|b\|\rangle)=O(1)$, then $\K(\|b\|)\lel 2\log\I(x:\ch)$.\\
\end{lmm}
\prf
	Due to Proposition \ref{prp:borderprefix}, by the definition of $b$, $b$ is total and $b^-$ is not, so ${b^-}\sqsubset\mathcal{B}$ is a prefix of border, and is thus a random string, with $\|b\|\lel\K(b)$. Due to Lemma~\ref{lmm:totalString}, with the second term removed,
	\begin{align*}
	\|b\|&\,{\lel}\,\K(b)\,{\lel}\, \I(x:\mathcal{H})+\K(b|\langle x,\|b\|\rangle)\\
	\|b\|&\lel \I(x:\mathcal{H})\\
	\K(\|b\|)&\lea 2\log\I(x:\mathcal{H}).
	\end{align*}\qed
\subsection{Stochasticity and the Halting Sequence}
Left-total machines can be used to prove properties of stochasticity. As mentioned earlier, the stochasticity of a string lower bounds the amount of mutual information it has with the halting sequence. The following lemma was first introduced in \cite{EpsteinLe11}.\\

\begin{figure}[h!]
	\begin{center}
		\includegraphics[width=0.7\columnwidth]{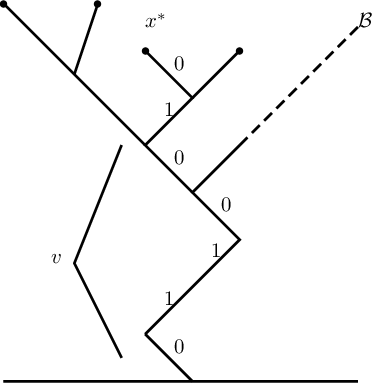}
		\caption{A graphical depiction of the terms used in Lemma \ref{lmm:StochH}. The shortest program for $x\in\N$ is $x^*=0110010$, with $U(x^*)=x$. The shortest total prefix of $x^*$ is $v=01100$, with $v^-=0110$ being a prefix of border $\mathcal{B}$. Assuming $x^*$ is the only extension of $v$ that is a program for $x$, then $Q(x)=2^{-\|x^*\|+\|v\|}=2^{-2}$.}
		\label{fig:StochHLemma}
	\end{center}
\end{figure}
\begin{lmm}
	\label{lmm:StochH}
	For  $x\in\N$, $\Ks(x)<\I(x\,{:}\,\ch)+6\K(\I(x\,{:}\,\ch))$.\\
\end{lmm}
\prf
	Using the optimal left-total Turing machine, let $U(x^*)=x$, $\|x^*\|=\K(x)$, and $v$ be the shortest total prefix of $x^*$. We define the elementary probability measure $Q$ such that $Q(a) = \sum_w 2^{-\|w\|}[U(vw)\,{=}\,a]$. A graphical depiction of these definitions can be seen in Figure \ref{fig:StochHLemma}. Thus $Q$ is computable relative to $v$. In addition, since $v\sqsubseteq x^*$, one has the lower bound $Q(x) \geq  2^{-\|x^*\|+\|v\|} = 2^{-\K(x)+\|v\|}$. Therefore 
	\begin{align}
	\nonumber
	\mathbf{d}(x|Q,v) &= \floor{-\log Q(x)} - \K(x|\langle Q\rangle, v) \\
	\nonumber
	&\eqa  -\log Q(x) - \K(x|v) \\
	\nonumber
	&\lea  \K(x)-\|v\| - \K(x|v)\\ 
	\nonumber
	&\lea (\K(v) +\K(x|v)) - \|v\| -\K(x|v)\\
	\nonumber
	&\lea (\|v\|+\K(\|v\|)+\K(x|v)) - \|v\| -\K(x|v)\\
	\mathbf{d}(x|Q,v) &\lea \K(\|v\|).\label{eqnDef}
	\end{align}
	
	Since $v$ is total and $v^{-}$ is not total, by  Proposition~\ref{prp:borderprefix}, $v^-$ is a prefix of the border sequence $\mathcal{B}$. In addition, $Q$ is computable from $v$. Therefore
	\begin{align}
	\nonumber
	\K(x|\ch) &\lea \K(x|Q) +\K(Q|\ch)\\
	\nonumber 
	&\lea \K(x|Q) + \K(v|\ch)\\
	&\lea -\log Q(x)+\K(\|v\|)\label{eq:Halting1}\\ 
	\nonumber
	&\lea \K(x) - \|v\| +\K(\|v\|)\\
	\nonumber
	\|v\| &\lea \K(x)-\K(x|\ch) +\K(\|v\|)\\
	\|v\| &\lea \I(x:\ch) + 2\K(\I(x:\ch)).\label{eq:Halting2}
	\end{align}
	
	Equation~(\ref{eq:Halting1}) is due to $\mathcal{B}$ being computable from $\mathcal{H}$, therefore $v^-\sqsubset\mathcal{B}$ is simple relative to  $\mathcal{H}$ and $\|v\|$. Equation~(\ref{eq:Halting2}) is from Lemma \ref{lmm:K}. Since $Q$ is computable from $v$, one gets 
	\begin{align*}
	\Ks(x) &\lea \K(v)+3\log(\max\{\mathbf{d}(x|Q,v),1\})\\
	&\lea \|v\|+\K(\|v\|)+3\log(\max\{\mathbf{d}(x|Q,v),1\})\\
	&\lea \|v\|+\K(\|v\|)+3\log \K(\|v\|)\\
	&\lea \|v\| +2\K(\|v\|).
	\end{align*}
	Applying Lemma \ref{lmm:abc} to $f\equiv\Ks(x)$, $g\equiv\|v\|$, and $h\equiv\I(x:\ch)$, with $n=2$ and $m=2$, gets $\Ks(x)\lea \I(x:\ch)+6\K(\I(x:\ch))$.\qed

\section{Discrete Sampling}
Theorem \ref{thr:stochsamp} has applications to sampling no-go theorems. In this section, we use this theorem to show that any sampling method will eventually produce outliers. The greater the sample size the greater the outlier score of an element in the sample. We first rework Theorem \ref{thr:stochsamp} to be in terms of mutual information with the halting sequence and not stochasticity.
\begin{cor}
	\label{thr:DisHalt}
	Relativized to computable semi-measure $P$ over $\N$, for any finite set $D\subset \N$, if $\N\ni s <\log\sum_{a\in D}\m(a)/P(a)$, then $s\lea\log \max_{a\in D}\m(a)/P(a)+\I(D:\ch)+6\K(\I(D:\ch))+2\K(s)$.
\end{cor}
\prf
	This follows from the application of Lemma \ref{lmm:StochH} to Theorem \ref{thr:stochsamp}.\qed

\begin{cor}
	\label{cor:DisHaltN}
	Relativized to computable semi-measure $P$ over $\N$, for any finite set $D\subset \N$, if $n< \log |D|$, $n\lea\log \max_{a\in D}\m(a)/P(a)+\I(D:\ch)+6\K(\I(D:\ch))+2\K(n)$.
\end{cor}
We recall that for a semi-measure $P$ over $\N$, a sampling method $A$ is a total computable function that takes in a parameter $n$ and a random source of bits and outputs, with probability one, an encoding of $2^n$ unique natural numbers.
\begin{cor}
	\label{cor:disprobsamp}$ $\\
	For  computable semi-measure $P$ over $\N$, for sampling method $A$, there is a constant $c_{P,A}\in\N$, where for all $n,k\in\N$,
	$\Pr(n-\log\max_{a\in A(n)}\m(a)/P(a)>k)<2^{-k+O(\K(n,k))+c_{P,A}}$. 
\end{cor}
\prf
	Given a fixed $n$ and $A$, let $X=\{x_i\}$ be the (possibly infinite) prefix free set of finite sequences representing the random seeds that cause $A$ to halt, with for each $i$, $A(n,x_i)=y_i$, where $y_i\in\FS$ is an encoding of $2^n$ natural numbers. Let $X_c\subseteq X$ be the subset of $X$ such that for all $x\in X_c$, $A(n,x)=y$, and $\I(y:\mathcal{H})\geq c$. Let $c_A\in\N$ be the size of a program that takes in $x_i$ and a program for $n$ and uses $A$ to output $y_i$. Thus $c_A$ is a constant solely dependent on $A$ and the universal Turing machine $U$. Let $d\in\R_{\geq 0}$ be defined by $2^d=\sup_x\m(x)2^{\K(x)}$.  Then over all $c$, $\sum_{x\in X_c}2^{-\|x\|}\leq 2^{-c+\K(n)+c_A+d}$. Otherwise there is a $c$ where, 
	\begin{align*}
	2^d &< \sum_{x_i\in X_c}2^{-\|x_i\|-\K(n)-c_A}2^c\\
	&\leq\sum_{x_i\in X_c}2^{-\|x_i\|-\K(n)-c_A}2^{\I(y_i:\mathcal{H})}\\
	&\leq\sum_{y_i}\sum_{A(n, x_j)=y_i, x_j\in X_c}2^{-\|x_j\|-\K(n)-c_A}2^{\I(y_i:\mathcal{H})}\\
	&<\sum_{y_i}\m(y_i)2^{\I(y_i:\mathcal{H})}\leq\sum_{y_i}2^d2^{-\K(y_i|\mathcal{H})}\\
	&< 2^d,
	\end{align*}
	causing a contradiction. So for all $n$, 
	\begin{align}
	\nonumber
	&\;\;\;\;\Pr(n - \log \max_{a\in A(n)}\m(a)/P(a)>k)\\
	\label{eq:probdis1}
	&<\Pr(c_P{+}\I(A(n):\mathcal{H}){+}6\K(\I(A(n):\ch)){+}2\K(n)\gea k)\\
	\label{eq:lognlogk}
	&<\Pr(c_P{+}\I(A(n):\mathcal{H}){\gea} k{-}2\K(n){-}12\K(k{-}2\K(n)))\\
	\nonumber
	&<\Pr(c_P+\I(A(n):\mathcal{H})\gea k-O(\K(n)+\K(k,\K(n))))\\
	\nonumber
	&<\Pr(\I(A(n):\mathcal{H})\gea k-O(\K(k,n))-c_P)\\	
	\nonumber
	&\leq\sum\left\{2^{-\|x\|}: x\in X_{k-O(\K(k,n))-c_P}\right\}\\
	\nonumber
	&< 2^{-k+O(\K(k,n))+c_A+c_P}.
	\end{align}
	
	Equation \ref{eq:probdis1} comes from the application of Corollary  \ref{cor:DisHaltN}. Equation \ref{eq:lognlogk} comes from Lemma \ref{lmm:aKgb}. The term $c_P\in\N$ is a constant solely dependent on $P$ and the universal Turing machine $U$. \qed

\section{Infinite Sequences}
\label{sec:DefInfSec}
In Section \ref{sec:DefRand}, the deficiency of randomness, $\d$, of natural numbers was defined. In this section, we define the deficiency of randomness{\tiny } $\D$ of infinite sequence. This notion will be used in the no-go sampling theorems over infinite sequences. Before introducing $\D$, we review some standard notions of measures and integration.

A set of subsets of a set $X$ is called an algebra if it is closed under finite intersections and unions and under complements. It is called a $\sigma$-algebra if it is closed under countable intersections and unions and under complements. 

A nonnegative function $\mu$ defined over some subsets of $X$ is monotonic if $A\subseteq B$ implies $\mu(A)\subseteq\mu(B)$. Such a function is additive if whenever $\mu$ is defined on disjoint $E_1,\dots,E_n$, then $\mu$ is defined on $E=\bigcup_{i=1}^nE_i$, and $\mu(E)=\sum_{i=1}^n\mu(E_i)$. It is $\sigma$-additive if whenever $\mu$ is defined on disjoint $\{E_i\}_{i=1}^\infty$, then $\mu$ is defined on $E=\bigcup_{i=1}^\infty E_i$, and $\mu(E)=\sum_{i=1}^\infty\mu(E_i)$. 

A pair $(X,\mathcal{A})$ consisting of a set $X$ and a $\sigma$-algebra $\mathcal{A}$ over $X$ is a measurable space. A measure $\mu$ is nonnegative $\sigma$-additive function over $\mathcal{A}$. It is a probability measure if $\mu(X)=1$. The triplet $(X,\mathcal{A},\mu)$ is called a measure space.

For this paper, we focus our attention on, $\IS$, the set of infinite sequences. For a string $x\in\FS$, the set of all infinite sequences that start with $x$, denoted $x\IS$, is called a cylinder set. For infinite strings, measures can be derived by functions on strings, $\mu:\FS\rightarrow \R_{\geq 0}$, where $\mu(x)=\mu(x0)+\mu(x1)$. Such functions are also referred to as measures. This is because $\mu$ can be defined on cylinder sets in the standard way, and then by the Carath\'{e}odory's extension theorem, to all Borel sets $\mathcal{B}$ of infinite sequences, which is the smallest $\sigma$-algebra containing the cylinder sets. Thus $(\IS,\mathcal{B},\mu)$ defines a measure space. Such measures $\mu$ are called probability measures if $\mu(\IS)=1$.  A measure $\mu:\FS\rightarrow\R_{\geq 0}$ is computable if it computable as defined in Section \ref{sec:conv}. 

Another example of a measurable space is $(\R,\mathcal{C})$ where $\mathcal{C}$ are the Borel sets of $\R$, i.e. the smallest $\sigma$ algebra containing the open intervals $(a,b)\subset\R$. We say a function $f:\IS\rightarrow\R$ is measurable if and only if $f^{-1}(C)\in\mathcal{B}$ whenever $C\in\mathcal{C}$. We say $f$ is continuous if for every $x\in\IS$, for every $\epsilon>0$, there is a cylinder set $Z\ni x$, such that $|f(x)-f(z)|<\epsilon$ for every $z\in Z$. A function $f:\IS\rightarrow\R$ is lower semi-continuous  if for every $r\in \R$, the set $\{x
\in \IS:f(x)>r\}$ is open. All lower semi-computable functions are by definition, lower semi-continuous.

A measurable function $g$ is simple if its range is finite: $\{a_1,\dots,a_k\}$. The (Lebesgue) integral of such $g$ is $\int gd\mu =\sum_{i=1}^ka_i\mu(g^{-1}(a_i))$. The integral of a measurable function $f$, is $\int fd\mu = \sup\{\int gd\mu:g\leq f,g\textrm{ is simple}\}$. A function $D:\IS\rightarrow\R_{\geq 0}\cup\infty$ is an integrable test with respect to computable probability measure $P$ if it is lower semi-computable and $\int_{\IS}2^{D(\alpha)}P(d\alpha)\leq 1$.\\

\noindent\textit{Theorem.} (\cite{Gacs21})
\textit{For computable probability measure $P$ over $\IS$, there exists a universal integrable test $\D:\IS\rightarrow\R_{\geq 0}\cup\infty$, where for all other integrable tests $D$,}
\begin{align*}
D(\alpha) &\lea \D(\alpha|P)+\K(D|P).
\end{align*}

As shown in the following theorem, any such universal integrable test $\D$ is equal, up to an additive constant, to a supremum of a term that uses the finite prefix of an infinite sequence.\\

\noindent\textit{Theorem.} (\cite{Gacs21})
\textit{For universal integrable test $\D$ for computable probability measure $P$ over $\IS$, 
	\begin{align*}
	\D(\alpha|P)\eqa \sup_{n\in \N} -\log P(\alpha[0..n])-\K(\alpha[0..n]|P),
	\end{align*}
	where the constant depends on $P$.}\\

This justifies the following definition.\\
\begin{dff}[Deficiency of Randomness of an Infinite Sequence]
	$\D(\alpha|P)=\sup_{n\in\N} -\log P(\alpha[0..n])-\K(\alpha[0..n]|P)$.\\ 
\end{dff}
As we look at sampling with respect to infinite sequences, we will need an information function between infinite sequences, and more specifically the amount of information that a specific sequence $\alpha$ has with the halting sequence $\mathcal{H}$. We use the symmetric function $\I:\IS\times\IS\rightarrow \R$, where\\
\begin{dff}[Information of Infinite Sequences]
	\label{dff:ii}\\
	For $\alpha,\beta\in\IS$, and $c\in\FS$,\\
	$\I(\alpha:\beta|c)=\log\sum_{x,y\in\FS}\m(x|c,\alpha)\m(y|c,\beta)2^{\I(x:y|c)}.$\\
\end{dff}
This function was introduced in \cite{Levin74}. The following theorem was stated in \cite{Levin74}, and a proof of it can be found in \cite{Vereshchagin21}.\\

\begin{thr}
	Assume that a family $P_\rho$, $\rho\in \Omega$, of probability distributions on $\Omega$ is fixed. Assume that there is a Turing machine $T$ that for all $\rho$ computes $P_\rho$ having oracle access to $\rho$. Then for all $\alpha$, $\rho \in\Omega$, there is a probability bounded (and even expectation) $P_\rho$-test $t_{\alpha,\rho,T}$ such that
	$$
	\I(\langle \rho,\omega\rangle:\alpha)\leq \I(\rho:\alpha)+t_{\alpha,\rho,T}(\omega)+c_T,
	$$
	for all $\omega\in\Omega$, where $c_T$ does not depend on $\rho$, $\alpha$, $\omega$.
\end{thr}
In \cite{Geiger12}, it is shown that the above theorem implies the following.\\

\begin{thr}
	\label{thr:IProbCons}
	Let $(P_\alpha)_{\alpha\in\IS}$ be a family of uniformly $\alpha$-computable continuous probability measures. Then for all $\alpha, \beta\in\IS$ we have 
	$$
	P_\alpha(\{\gamma\in\IS:\I(\langle \alpha,\gamma\rangle:\beta)-\I(\alpha:\beta)>m\}) \leq 2^{-m+c_{\alpha,\beta}},
	$$
	where $c_{\alpha,\beta}$ is a positive constant dependent solely on $\alpha$ and $\beta$.
\end{thr}

In addition \cite{Geiger12} contains a short proof for the following theorem.\\
\begin{thr}
	\label{thr:detconsi}
	For partial recursive $f:\IS\rightarrow\IS$, $\alpha,\beta\in \IS$,
	$\I(f(\alpha):\beta)\lea \I(\alpha:\beta)+\K(f)$.
\end{thr}
\section{Continuous Sampling}

This section proves sampling no-go theorems for infinite sequences. Theorem \ref{thr:InfExt} uses the following definitions. We recall that $x\sqsubseteq y$ for $x,y\in\FS$ implies that $x$ is a prefix of $y$ or equal to $y$. For a string $x\in\FS$, let 	$\D(x|P)= \max_{y\sqsubseteq x}(\log (\m(y|P)/P(y)))$. Let $\B(b)=\max\{U(p):p\lhd b,\textrm{ or }p\sqsupseteq b\}$ be the largest number produced by a program that extends $b$ or is to the left of $b$.\\
\label{sec:Cont}
\begin{thr} 
\label{thr:InfExt} 
Relativized to computable probability measure $P$ over $\IS$, for $Z\subseteq \IS$, if $\N\ni s< \log\sum_{\alpha\in Z}2^{\D(\alpha|P)}$, then $s<\sup_{\alpha\in Z}\D(\alpha|P)\,{+}\,\I(\langle Z\rangle:\ch)+O(\K(s)+\log\I(\langle Z\rangle:\ch))$.
\end{thr}$ $\\
\noindent\textit{Informal Proof.} The proof starts off by determining an $N\in\N$, such that $\sum_{x\in Z_{\leq N}}2^{\D(x|P)}>2^s$. This $N$ is equal to $\B(b)$ for some total string $b$. Then Lemma \ref{lmm:main} is invoked with $W(x)=P(x)$, $\eta(x)=[x\in\BT^N]2^{\D(x|P)}$, $D=Z_{\leq N}$, relativized to $b$. This produces $x\in D$ where $\K(x|b) \lea -\log P(x) + \Ks(D|b) + O(\K(s))$. Using Lemma \ref{lmm:StochH}, the $\Ks(D|b)$ term is replaced with $\I(D:\ch|b)$. The conditioning on $b$ is removed using Lemma \ref{lmm:totalString}. Finally the $\I(D:\ch)$ term is replaced with $\I(\langle Z\rangle:\ch)$ to achieve the theorem.

\prf
$ $\\ \\
\noindent \textit{1. Determination of $N$.}\\
For a total $b\in\FS$, let $\m_b(x|y)=\sum\{2^{-\|z\|}:U_y(z)=x,\, U_y(z)\textrm{ halts in }\bb(b)\textrm{ time}\}$ be the algorithmic weight of $x$ using solely programs that are running in $\bb(b)$ time. For $x\,{\in}\,\FS$, let $\D_b(x|P)= \max_{y\sqsubseteq x}(\log (\m_b(y|P)/P(y)))$, with $\D_b\leq \D$. We set $b$ to be the shortest total string with
\begin{enumerate}
	\item $N=\B(b)$.
	\item $\sum_{x\in Z_{\leq N}}2^{\D_b(x|P)}{>} 2^s$.\\
\end{enumerate}

\noindent\textit{2. Invocation of Lemma \ref{lmm:main}.}\\
We let $W(x)= P(x)$, $\eta(x)= 2^{\D(x|P)}[x\,{\in}\,\BT^N]$, and  $D= Z_{\leq N}$. Since the universal Turing machine is relativized to $P$, it must be that $\K(\langle W,\eta\rangle| b)=O(1)$, $\log\sum_{x\in D}\eta(x)> s$, and
\begin{align*}
\sum_{x\in\FS}W(x)\eta(x) & =\sum_{x\in\BT^N}P(x)2^{\D(x|P)}                    \\
& = \int_{\alpha}2^{\D(\alpha[0..N]|P)}dP(\alpha) \\
& \leq\int_{\alpha}2^{\D(\alpha|P)}dP(\alpha){\leq} 1.
\end{align*}
Lemma~\ref{lmm:main}, relativized to $b$, gives $x\,{\in}\, D$ with 
\begin{align*}
\K(x|b)&< -\log P(x)-s +\Ks(D| b)+O(\K(s)).
\end{align*}
\noindent\textit{3. Replace $\Ks(D|b)$ with $\I(D:\ch|b)$.}\\
Due to Lemma \ref{lmm:StochH}, 
\begin{align*}
\K(x|b)&< -\log P(x)-s+
\I(D:\ch|b)\\
&\;\;\;\;+O(\K(s)+\log\I(D:\ch|b))\\
s&< \log (\m(x)/P(x))+\K(b)+\I(D:\ch|b)\\
&\;\;\;\;+O(\K(s)+\log (\I(D:\ch|b)+\K(b))).
\end{align*} 
\noindent\textit{4. Remove conditioning of $b$.}\\
By Lemma~\ref{lmm:totalString}, 
\begin{align*}
\K(b)+\I(D:\ch|b)&\lel \I(D:\ch)+\K(b|\langle D,\|b\|\rangle).
\end{align*} 
Therefore 
\begin{align*}
s&\leq \log (\m(x)/P(x))+\I(D:\ch)+\K(b|\langle D,\|b\|\rangle)\\
&\;\;\;\;+O(\K(s)+\log(\I(D:\ch)+\K(b|\langle D,\|b\|\rangle))).
\end{align*}

Since $D\subseteq {\BT}^{\bb(b)}$, $\K(b| \langle D,\|b\|\rangle)\,{=}\,O(1)$, as a program can output the leftmost total string $y$ of length $\|b\|$ such that $\bb(y)$ is the length of the strings in $D$. Hence 
\begin{align*}
s&\leq \log (\m(x)/P(x))+\I(D:\ch)+O(\K(s)+\log \I(D:\ch)).
\end{align*}
\noindent\textit{5. Replace $\I(D:\ch)$ with $\I(\langle Z\rangle:
	\ch)$.}\\
We have that $\K(D|\langle Z\rangle)\lea \K(\|b\|)+\K(s)$, as $D$ is computable from $\langle Z\rangle$, $\|b\|$, and $s$. This is because $b$ is computable from $\|b\|$, $s$, and $\langle Z\rangle$ and thus so is $D=Z_{\leq \bb(b)}$. By Definition \ref{dff:ii} of mutual information between infinite sequences,
\begin{align}
\nonumber
\I(D:\ch)&\lea \I(\langle Z\rangle:\ch)+\K(D|
\langle Z\rangle)\\
\nonumber
&\lea \I(\langle Z\rangle:\ch)+\K(\|b\|)+\K(s)\\
\label{eq:ZHs}
&\lea \I(\langle Z\rangle:\ch)+2\log \I(D:\ch)+\K(s)\\
\label{eq:DZs}
&\lel  \I(\langle Z\rangle:\ch)+\K(s). 
\end{align}
Where Equation \ref{eq:ZHs} is due to the application of Lemma \ref{lmm:KTotalLength}, noting $\K(b|\langle D,\|b\|\rangle)=O(1)$. Equation \ref{eq:DZs} is due to Lemma \ref{lmm:log}. So
\begin{align*} 
s&\leq \log (\m(x)/P(x)) +\I(D:\mathcal{H})+ O(\K(s)+\log\I(D:\mathcal{H}))\\
&\leq \sup_{\alpha\in Z}\D(\alpha |P)+\I(\langle Z\rangle:\ch)+O(\K(s)+\log\I(\langle Z\rangle:\ch)).
\end{align*} \qed\\

\begin{cor}
\label{cor:contn}
Relativized to computable probability measure $P$ over $\IS$, for any set $Z\subseteq\IS$ with $n<\log |Z|$, $n<\sup_{\alpha\in Z}\D(\alpha|P)+\I(\langle Z\rangle:\ch)+O(\K(n)+\log\I(\langle Z\rangle:\ch))$.
\end{cor}
\prf
This follows from the fact that for any $\alpha\in\IS$, $\D(\alpha|P)\gea 0$ because using continuous Shannon-Fano encoding, there is a prefix $x\sqsubset\alpha$ that can be identified by a code of $\eqa-\log P(x)$. This implies $\K(x|P)\lea -\log P(x)$ and thus $\D(\alpha|P)\geq -\log P(x)-\K(x|P)\gea 0$. Therefore there is some $c\in\N$ solely dependent on the universal Turing machine $U$, such that $\log\sum_{\alpha\in Z}2^{\D(\alpha|P)}>\log\sum_{\alpha\in Z}2^{-c}>n-c$.\qed\\

A continuous sampling method $A$ takes in a parameter $n$, a infinite source of random bits and outputs $2^n$ unique infinite sequences encoded in the form $$\alpha_1[1]\alpha_2[1]...\alpha_{2^n}[1]\alpha_1[2]\alpha_2[2]...\alpha_{2^n}[2]\dots$$ We get the following continuous sampling corollary which is analogous the discrete case.\\

\begin{cor}
\label{cor:contprobsamp}
For computable measure $P$ over $\IS$, for continuous sampling method $A$, there  exists $c_{P,A}\in\N$, where for all $n,k\in\N$,
$\Pr(n-\max_{\alpha\in A(n)}\D(\alpha|P)>k)<2^{-k+O(\log k+\K(n))+c_{P,A}}$.
\end{cor}
\prf
We use $\gamma\sim\mathcal{U}$ to represent infinite sequences distributed according to the uniform distribution.
\begin{align}
\nonumber
&\Pr_{\gamma\sim\mathcal{U}}
\left(n-\max_{\alpha\in A(n,\gamma)}\D(\alpha|P) > k\right)\\
\label{eq:contsamp1}
&<\Pr_{\gamma\sim\mathcal{U}}(c_P+\I(A(n,\gamma):\ch)+O(\log\I(A(n,\gamma):\ch) )\\
&\;\;\;\;\;\;\;\;\;\;\;\;>k-O(\K(n)))\\
\label{eq:contsamp2}
&<\Pr_{\gamma\sim\mathcal{U}}(c_P+\I(A(n,\gamma):\ch)>k-O(\K(n)+\log k))\\
\label{eq:contsamp3}
&<\Pr_{\gamma\sim\mathcal{U}}(\I(\gamma:\ch)>k-O(\K(n)+\log k)-c_P-c_A)\\
\label{eq:contsamp4}
&<2^{-k+O(\log k+\K(n))+c_{P}+c_{A}}.
\end{align}
Equation \ref{eq:contsamp1} comes from Corollary \ref{cor:contn}, where $c_P\in\N$ is a constant solely dependent on $P$ and the universal Turing machine $U$. Equation \ref{eq:contsamp2} comes from the fact that $a < i+O(\log i)$ implies that either $a<i$ or then $O(\log a)>O(\log i)$ and then $a-O(\log a)<i$. Equation \ref{eq:contsamp3} comes from Theorem \ref{thr:detconsi}, where $f=A(n,\cdot)$, with $\K(f)=O(\K(n))$. Thus $c_A\in\N$ is a constant solely dependent on $A$ and the universal Turing machine $U$. Equation \ref{eq:contsamp4} comes from the application of Theorem \ref{thr:IProbCons}, where $\alpha=0^\infty$, $\beta=\ch$, and $P_\alpha=\mathcal{U}$.\qed

\section{Completing Binary Predicates}
\label{sec:ComplPred}

A binary predicate is defined to be a function of the form $f:D\rightarrow\BT$, where $D\subseteq\N$.  We say that binary predicate (or finite string) $\lambda$ is an extension of $\gamma$, if  for all $i\in \mathrm{Dom}(\gamma)$, $\gamma(i)=\lambda(i)$. If a binary predicate has a domain of $\N$ and is an extension of binary predicate $\gamma$, then we say it is a complete extension of $\gamma$. The self-delimiting code for a binary predicate $\gamma$ with a finite domain is $\langle\{x_1,\lambda(x_1),\dots,x_n,\lambda(x_n)\}\rangle$. The Kolmogorov complexity of a binary predicate $\lambda$ with an infinite sized domain is $\K(\lambda)=\K(f)$, where $f:\N\rightarrow\N$ is a partial computable function where $f(i)=\lambda(i)$ if $i\in\mathrm{Dom}(\lambda)$ and $f(i)$ is undefined otherwise.  If there is no such partial computable function, then $\K(\lambda)=\infty$. 
\begin{thr} 
	\label{thr:PredExt}
	\noindent For binary predicate $\gamma$ and the set $\Gamma$ of complete extensions of $\gamma$, $\min_{g\,\in\,\Gamma}\K(g)\,{\lel}\,|\mathrm{Dom}(\gamma)|{+}\I(\langle\gamma\rangle{:}\mathcal{H}).$\\
\end{thr}

\prf
	We recall that $\B(b)=\max\{U(p):p\lhd b,\textrm{ or }p\sqsupseteq b\}$ is the largest number produced by a program that extends or is to the left of $b$. The theorem is meaningless if $|\mathrm{Dom}(\gamma)|=\infty$, so we can assume $q=|\mathrm{Dom}(\gamma)|<\infty$. Let $n=\max\{i:i\in\mathrm{Dom}(\gamma)\}$. Let $b$ be the shortest total string where $\B(b)\geq n$. Let $N=\B(b)$. It must be that $\K(b|\langle \gamma\rangle,\|b\|)=O(1)$ as there is a program that can enumerate, from the left, total strings of length $\|b\|$. This program returns the first total string $b'$ such that $\bb(b')\geq n$. This $b'$ is equal to $b$, otherwise $b'\lhd b$ and thus $\bb({b'}^-)\geq \bb(b')\geq n$, contradicting the definition of $b$.
	
	Let $D$ be the set of all strings of length $N$, that extends $\gamma$.  Lemma \ref{lmm:main}, relative to $b$, with $W(a)=1$, and $\eta(a)=[\|a\|=N]2^{-N}$, $s=\log \sum_{a\in D}\eta(a)=-q$ results in $a\in D$, with
	\begin{align}
	\label{eq:PredStoch}
	\K(a|b) & \lea q + \Ks(D|b)+2\K(q).
	\end{align}
	Lemma \ref{lmm:StochH} applied to Equation \ref{eq:PredStoch}, results in
	\begin{align*}
	\K(a|b) &\lel q + \I(D:\ch|b).
	\end{align*}
	Since $\K(D|b)\lea \K(\gamma|b)$ and $\K(\gamma|b,\ch)\lea \K(D|b,\ch)$,
	\begin{align}
	\nonumber
	\K(a|b) &\lel q +\I(\gamma:\ch|b)\\
	\label{eq:PredH}
	\K(a) &\lel q +\K(b)+\I(\gamma:\ch|b).
	\end{align}
	Lemma \ref{lmm:totalString}, applied to Equation \ref{eq:PredH}, results in
	\begin{align}
	\nonumber
	\K(a) &\lel q + \I(\gamma:\ch)+\K(b|\langle \gamma\rangle,\|b\|)\\
	\label{eq:PredF}
	\K(a) &\lel |\mathrm{Dom}(\gamma)|+\I(\gamma:\ch).
	\end{align}
	Thus there exists a complete extension $g'\in \Gamma$, of $\gamma$, that is equal to $a[i]$ for all $i\leq \|a\|$, and 0 otherwise. This $g'$ can be computed with a program of size $\lea \K(a)$, thus combined with Equation \ref{eq:PredF},
	\begin{align*}
	\min_{g\in \Gamma}\K(g)\leq \K(g')\lea \K(a) \lel |\mathrm{Dom}(\gamma)|+\I( \gamma:\ch).
	\end{align*}\qed 

\section{Discussion}
One area of progress is to improve the bounds in Corollary  \ref{cor:contprobsamp} to match that of the discrete case. There are several extensions or variants that can be made to the results in this paper. One is to replicate the result on deficiencies of randomness with respect probability measures over general spaces. In \cite{Epstein20}, a variant to Theorem \ref{thr:stochsamp} was used to provide new bounds between different algorithmic quantum entropies, one introduced in \cite{Vitanyi00}, and the other in \cite{Gacs01}. By leveraging the work in  \cite{Romashchenko03}, a conditional complexity alternative to \cite{EpsteinLe11} can be proven, that shows all natural sets of strings contain members that are simple to all its other members. In general, there are many ways of leveraging stochasticity to reason about combinatorial objects that are created by randomized methods.

%





\end{document}